\documentclass{article}                                                                  
\usepackage{amsmath}

\begin{document}

\title{Topics  in  Born-Infeld  Electrodynamics }                                         
                                                                                          
\author{R. Kerner, 
A.L. Barbosa\thanks{Permanent address: IFT,     
Universidade Estadual Paulista, Rua Pamplona 145, 
 01405 São Paulo, Brazil.}                        
\ and \
D.V. Gal'tsov\thanks{Permanent address: Department of                
 Theoretical Physics, Moscow State University, 119899 Moscow, Russia.} 
\\ \\                                                                        
L.P.T.L. - Tour 22, 4-\`{e}me \'{e}tage,  \\                                      
 Bo{i}te 142, Universit\'{e} Paris-VI, 
\\
\, 4, Place Jussieu, 75005 Paris}

\maketitle

\section{Introduction - a Short Glance at the History} 

 Since the discovery of the electron by
J.J.Thomson in 1899, physicists tried to develop models of
finite energy charge concentrations that could describe the
elementary electric charge. One of the ideas was to use
non-linear generalizations of Maxwell's theory, deviating from
it only at very short distances and very strong fields in order
to ensure a cut-off and to avoid singularity at $r \rightarrow
0$.
\newline
\indent G. Mie (\cite{Mie}) was first to introduce such a
model, based on the assumption that the electric field ${\bf
E}$ can not exceed the limiting value ${\bf E_0}$, and that the
repulsive force should be proportional to the expression
\begin{equation}
{\bf F} \sim \frac{\bf E}{\sqrt{1 - \frac{{\bf E}^2}{\bf
E_0^2}}} \, . \label{kern-eq1.1}
\end{equation}
In this model it was possible to find a nonsingular solution
with finite energy and charge, and with the field ${\bf E}$
falling off as $r^{-2}$ at great distances, but this solution
was not covariant with respect to the Lorentz transformations.
\newline
\indent In 1932 and in 1934 Born and Infeld have published by
now celebrated version of non-linear electrodynamics, in which
they proposed the following Lorentz-invariant Lagrangian:
\begin{equation}
{\cal{L}} =  \beta^2 \biggl[ \, \sqrt{{\rm det} \, \biggl(
\delta^{\mu}_{\lambda} + {\beta}^{-1} \, F^{\mu}_{\, \, \lambda}
\biggr) } \biggr] \, . \label{kern-eq.1.2}
\end{equation}
\indent The constant $\beta$ appears for dimensional reasons,
and plays the same r\^ole here as the limiting value of the
electric field in G. Mie's non-linear electrodynamics.
\newline
\indent When expressed in terms of two invariants of Maxwell's
tensor,
$$P = \frac{1}{4} \, F_{\mu \nu} \, F^{\mu \nu}  \, \ \ {\rm and}
\, \ \ \, \ \ S = \frac{1}{4} \, F_{\mu \nu} \,
{\tilde{F}}^{\mu \nu} \, ,
 \ \ {\rm with} \, \ \ \, {\tilde{F}}^{\mu \nu} = \frac{1}{2} \,
 \, \epsilon^{\mu \nu \lambda \rho } \, F_{\lambda \rho} $$
this Lagrangian can be written explicitly as
\begin{equation}
{\cal{L}}_{BI}  = \beta^2 \, \biggl[ 1 - \sqrt{1 + 2 P - S^2}
\biggr] \, , \label{kern-eq.1.3}
\end{equation}
\begin{equation}
\, \ \ {\rm or \ \ as} \ \ \, \ \ {\cal{L}}_{BI} = \beta^2 \,
\Biggl[ \, \ \
 1 - \sqrt{ 1 + \frac{1}{2 \beta^2} \, ({\bf B}^2 - {\bf E}^2 ) -
\frac{1}{16 \beta^4} \, ( {\bf E} \cdot {\bf B} )^2 } \, \ \ \,
\Biggr] \, . \label{kern-eq.1.4}
\end{equation}
\indent With the advent of Quantum Mechanics and Dirac's
equation for the electron, the interest in classical models of
charged particles has considerably faded. However, in 1970 G.
Boillat (\cite{Boillat} considered the Born-Infeld
electrodynamics as an example of non-linear theory in order to
study its propagation properties. Starting with the most
general non-linear theory derived from an arbitrary Lagrangian
depending on two Lorentz invariants of Maxwell's tensor,
${\cal{L}} \, (P,S)$, he discovered that among all such
non-linear theories, the Born-Infeld electrodynamics is the
only one ensuring the absence of birefringence, i.e.
propagation along a single light-cone, and the absence of shock
waves. In this respect the Born-Infeld theory is unique (except
for another singular and unphysical Lagrangian ${\cal{L}} =
P/S$). A beautiful discussion of these properties can be found
in I. Bialynicki-Birula's paper (\cite{Birula}); an interesting
generalization of Born-Infeld theory in a curved space-time 
background can be found in the paper by L.N. Chang {\it et al},
(\cite{Mansouri}). Let us remind very shortly how the 
birefringence phenomenon may occur in non-linear theories.
\newline
\indent From the mathematical point of view, these theories are
based on systems of second order partial differential
equations, linear in highest derivatives, with coefficients
which depend only on the fields (but not on their derivatives).
The systems of this type can be reduced to a set of differential
equations of first order via introduction of auxiliary fields,
which are the independent linear combinations of the first
partial derivatives of functions corresponding to the degrees
of freedom of our system.
\newline
\indent The differential system can be represented by means of
a matrix whose entries contain the operators of partial
derivation or multiplicative coefficients, acting on a
vector-column representing auxiliary fields. If the
vector-column ${\bf u}$ with the fields $\psi$, $\chi_i$, $E_i$
and $B_i$ contains $N$ elements, then let us denote by ${\cal
A}$ the $N$x$N$ matrix containing the partial derivatives and
by ${\cal B}$ the $N$x$N$ matrix containing the multiplicative
factors. Then the field equations can be written as:
\begin{equation}
{\cal A}^{\mu} ({\bf u}) \partial_\mu {\bf u} + {\cal B}({\bf
u}) {\bf u} = 0  \, . \label{kern-eq.1.5}
\end{equation}
If the hypersurface defined by the implicit equation
\begin{equation}
\Sigma \, (x^{\mu})  \, = 0  \label{kern-eq.1.6}
\end{equation}
is a surface of discontinuity, then the first derivatives of
fields are discontinuous across this surface, whereas the
fields themselves are continuous. So, when applied to the {\it
discontinuities} across the hypersurface (\ref{kern-eq.1.6}),
the equation (\ref{kern-eq.1.5}) reduces to
\begin{equation}
\left({\cal A}^{\mu} \, \Sigma_{\mu} \right) \delta_1 {\bf u} =
0 \, , \label{kern-eq.1.7}
\end{equation}
where $\Sigma_{\mu} \equiv \partial_\mu \Sigma$, and $\delta_1
{\bf u}$ denotes the discontinuity of the first derivative
across $\Sigma$, $\delta_1 {\bf u} \equiv \partial {\bf
u}/\partial\Sigma|_+ - \partial {\bf u}/\partial\Sigma|_-$.  By
definition, for a characteristic surface one has $\delta_1 {\bf
u}\ne 0$, therefore, in order for (\ref{kern-eq.1.7}) to hold,
one must have
\begin{equation}
{\rm det} \biggl( {\cal A}^{\mu} \, \Sigma_{\mu} \biggr) = 0 \,
, \label{kern-eq.1.8}
\end{equation}
on the surface of discontinuity. The characteristic equation
(\ref{kern-eq.1.8}) determines the surface whose generic
equation is $H(x,\Sigma_\mu)=0$, with $H$ a homogeneous
function of order $N$ in $\Sigma_\mu$. The Born-Infeld theory
turns out to be {\it completely exceptional} since it obeys the
corresponding condition of \cite{Boillat}, namely $\delta_0
H\equiv H|_+ - H|_- =0.$
\newline
\indent Let us give an illustration of this principle on the
simplest case: the scalar field wave equation in a
two-dimensional space-time $(t,x)$:
\begin{equation}
\partial^2_0 \, \phi - \partial^2_x \, \phi = 0 \, .
\label{kern-eq.1.9}
\end{equation}
(Partial derivatives in Lorentz indices, $(0,x,y,z)$ or
$(0,1,2,3)$, will be denoted by $\partial_0, \, \partial_x$),
etc..  According to the prescription, we can use as auxiliary
fields $\psi$ and $\chi$ the first derivatives of the scalar
field $\phi$, $\partial_0 \phi = \psi $ and  $\partial_x \phi =
\chi$. Then by definition, the first derivatives of auxiliary
fields are not independent, because we have, as $\partial_0 \,
(\partial_x \, \phi) =
\partial_x \, (\partial_0 \, \phi)$, automatically $\partial_0
\chi - \partial_x \psi = 0$.  On the other hand the dynamical
equation (\ref{kern-eq.1.9}) can be written as $\partial_0 \psi
- \partial_x \chi = 0$. In the matrix notation of
(\ref{kern-eq.1.7}) these two equations can be combined to yield
\begin{equation}
\begin{pmatrix}
 0 & 1 \cr 1 & 0
\end{pmatrix}
\, \partial_0  
\begin{pmatrix}
 \psi \cr
\chi
\end{pmatrix}
 + 
\begin{pmatrix}
 -1 & 0 \cr   0 & -1 
\end{pmatrix}
 \, \partial_x
\begin{pmatrix}
\psi \cr \chi
\end{pmatrix}
 = 
\begin{pmatrix}
 0 \cr 0
\end{pmatrix}
\, .
\label{kern-eq.1.10}
\end{equation}
We then find
\begin{equation}
{\cal A}^\mu\Sigma_\mu = 
\begin{pmatrix}
 - \Sigma_x & \Sigma_0 \cr
\Sigma_0 & -\Sigma_x
\end{pmatrix}
\, , \label{kern-eq.1.11}
\end{equation}
and the characteristic equation ${\rm det}({\cal
A}^\mu\Sigma_\mu)=0$ can be written as
\begin{equation}
{\Sigma_0}^2 - {\Sigma_x}^2 = 0 \,  , \label{kern-eq.1.12}
\end{equation}
where $\Sigma_0\equiv\partial_0 \Sigma$ and
$\Sigma_x\equiv\partial_x \Sigma$. The last equation defines
the characteristic surfaces $\Sigma(t,x)$, which in this case
are the light-cones in two space-time dimensions.
\newline
\indent The same technique can be easily applied to the
electromagnetic Maxwellian field. We have to solve a $6 \times
6$ matrix, because we have now six independent combinations of
its first derivatives (the fields ${\bf E}$ and ${\bf B}$)
appearing in the first-order Maxwell's equations. As we know,
the characteristic surfaces in four dimensions are given by
$\Sigma_{,\mu} \, \Sigma^{,\mu} = {\Sigma_0}^2 - {\Sigma_x}^2
-{\Sigma_y}^2 - {\Sigma_z}^2 =0$.
\newline
\indent The same is true for the Born-Infeld non-linear
electrodynamics. A more exhaustive discussion of the
propagation properties of various non-linear generalizations of
the electromagnetism can be found in recent papers (
\cite{LemosKerner}, \cite{GibbonsHerdeiro}).
\newline
\indent An entirely new and unexpected impulse for the revival
of interest in the Born-Infeld electrodynamics, and in its
non-abelian generalizations, came from recent developments of
the string and brane theories. The string Lagrangian in (4 + D)
dimensions, which defines a minimal surface in a
(4+D)-dimensional Minkowskian space-time, is in fact a
generalization of geodesic equation for a point-like particle.
\newline
\indent Consider a two-dimensional surface with cylindrical
topology, parametrized with one time-like and one space-like
parameter, $\tau$ and $\sigma$, respectively, and embedded in a
(4+D)-dimensional space-time. The embedding functions will be
denoted by  $X^{\mu} \, (\tau, \sigma)$, or equivalently, as
$X^{\mu} \, (\xi^a)$, with $\mu, \nu = 0,1,2,...3+D$ , and
$A,B,..= 1,2$ so that \hbox{$\xi^1 = \tau, \, \xi^2 = \sigma$.}
\newline
\indent The exterior space-time metric $g_{\mu \lambda}$
induces the internal metric of the world-sheet spanned by the
string,
\begin{equation}
G_{AB} = g_{\mu \lambda} \, \partial_a X^{\mu} \, \partial_b
X^{\lambda} \, .  \label{kern-eq.1.13}
\end{equation}
Let $h^{AB}\, (\xi^c)$ be an arbitrary metric on the
world-sheet; the variational principle introduced first by A.
Polyakov reads then
\begin{equation}
\delta \, S = - \frac{1}{4 \pi {\alpha}' } \, \delta \, \int
\int \, \sqrt{- h} \, h^{AB} \, G_{AB} \, {\rm d} \tau {\rm d}
\sigma \, = 0  \, . \label{kern-eq.1.14}
\end{equation}
Under the independent variations $\delta x^{\mu}$ and $\delta
\, h^{AB}$ one gets the following equations:
\begin{eqnarray}
G_{AB} - \frac{1}{2} \, h_{AB} \biggl( h^{CD} \, G_{CD} \, \biggr) = 0 ,
\\
h^{AB} \, \Biggl[ \, \nabla_A \nabla_B x^{\mu} +
\Gamma^{\mu}_{\lambda \rho} \, \partial_A x^{\lambda}
\partial_B x^{\rho} \Biggr] = 0. \label{kern-eq.1.15-16}
\end{eqnarray}
\indent After dimensional reduction from 11 to 10 dimensions,
auxiliary fields $A_{\mu}$ and $\phi$ do appear, and the total
Lagrangian takes on the form that contains the Born-Infeld
Lagrangian (\cite{Ts97,GaGoTo98,BrPe98}).
\begin{equation}
{\cal{L}} = \frac{1}{2} \, {\cal{D}}_{\mu} \Phi {\cal{D}}^{\mu}
\Phi^* + \beta^2 \, (1-{\cal{R}}) - \frac{\lambda}{2} \, (
\Phi^* \, \Phi - v^2)^2 + \frac{1}{16 \pi G} \, R
\label{kern-eq.1.17}
\end{equation}
with $\Phi$ denoting scalar field, $R$ the Riemann curvature
scalar, and ${\cal{R}}$ given by
\begin{equation}
{\cal{R}}=\sqrt{1+\frac{1}{2\beta^2} F^a_{\mu\nu}F_a^{\mu\nu}
-\frac{1}{16\beta^4}(F^a_{\mu\nu}{\tilde F}_a^{\mu\nu})^2}\, .
\label{kern-eq.1.18}
\end{equation}
\indent For dimensional reductions onto lower dimensions, the
non-abelian generalizations of this Lagrangian are naturally
produced.
\newline
In a pure Yang-Mills theory in flat space-time, with the usual
Lagrangian density 
 \hbox{${\cal{L}}_{YM} = -\frac{1}{4} g_AB \,
 F^A_{\mu \nu} \, F^{B \, \mu \nu}$} 
 there are no finite energy
static non-singular solutions describing a charged soliton.
This fact can be explained by the conformal invariance of the
theory, and the tracelessness of the energy-momentum tensor,
\begin{equation}
T^{\mu}_{\, \ \ \mu} = -T_{00} +
{\displaystyle{\sum_{i=1}^{3}}} T_{ii} = 0\, . \label{kern-eq.1.19}
\end{equation}
Given the positivity of energy, i.e.$T_{00} > 0$, this means
that the sum of principal pressures is positive, too, $\sum
T_{ii} > 0$, which leads to the conclusion that the Yang-Mills
``matter'' is naturally subjected to repulsive forces only.
\newline
\indent The presence of the Higgs field breaks the conformal
invariance, which leads to the existence of 't Hooft and
Prasad-Sommerfield magnetic monopoles. In what follows, we are
interested in soliton-like solutions arising in other
non-linear theories, including non-abelian versions of
Yang-Mills theories, which are no more conformally invariant.

\section{Non-linear Electrodynamics from the
 Kaluza-Klein Theory}
 
 An interesting
non-linear generalization of electrodynamics derived from the
Kaluza-Klein theory in five dimensions has been proposed in
(\cite{Kerner1}, \cite{Kerner2}). It is based on the addition
of the Gauss-Bonnet term, $R_{A B C D} \, R^{A B C D} - 4 \,
R_{A B } \, R^{A B} + R^2$, which in five dimensions is not a
topological invariant, leading to non-trivial equations of
motion of second order when added to the Einstein-Hilbert
Lagrangian.
\newline
\indent In a flat space-time and without the scalar field the
Kaluza-Klein metric is
\begin{equation}
g_{AB} = 
\begin{pmatrix}
 g_{\mu \nu} + A_{\mu} A_{\nu} & A_{\mu} \cr
A_{\nu} & 1
\end{pmatrix}
 \, , \label{kern-eq.2.1}
\end{equation}
where $A,B=0,1,2,3,5$ and $\mu,\nu=0,1,2,3$ (or
$\mu,\nu=0,x,y,z$ following the convention we have been using).
The full Lagrangian is taken to be (see \cite{Kerner1,Kerner2}):
\begin{equation}
{\cal L} = R + \gamma \,( R_{A B C D} \, R^{A B C D} - 4 \,
R_{A B} \, R^{A B} + R^2 ) \, , \label{kern-eq.2.2}
\end{equation}
with $\gamma$ being a certain dimensional parameter
characterizing the strength of the non-linearity. When
expressed in four dimensions in terms of the Maxwell tensor, it
becomes
\begin{equation}
{\cal L} = - \frac{1}{4} F_{\mu \nu}\, F^{\mu \nu} - \frac{3
\gamma}{16} \, \Biggl[\, ( F_{\mu \nu} F^{\mu \nu} )^2  - 2 \,
(F_{\mu \lambda} F_{\nu \rho} F^{\mu \nu} F^{\lambda \rho}) \,
\biggr] \,  . \label{kern-eq.2.3}
\end{equation}
In terms of the invariants $P$ and $S$ this Lagrangian is given
by ${\cal L} = 2P + \frac{3\gamma}{2} S^2$, which for the
choice $\gamma=-\frac{2}{3}$ yields essentially the square of
the Born-Infeld Lagrangian. The equations of motion are:
\begin{equation}
F_{\lambda \rho,\mu} +  F_{\rho \mu,\lambda} + F_{\mu
\lambda,\rho} = 0 \, , \label{kern.eq.2.4}
\end{equation}
which correspond to the Bianchi identities and are geometrical
equations valid independently of the Lagrangian chosen, and the
dynamical equations resulting from the variational principle,
\begin{equation}
[ F^{\lambda \rho} - \frac{3 \gamma}{2} \, (F_{\mu \nu} F^{\mu
\nu}) F^{\lambda \rho} + \frac{3 \gamma}{2} F_{\mu \nu}
F^{\lambda \mu} F^{\rho \nu} ]_{,\lambda}= 0 \, .
\label{kern-eq.2.5}
\end{equation}
The Lagrangian (\ref{kern-eq.2.3}) is particularly simple when
expressed in more familiar terms with the fields $\bf E$ and
$\bf B$:
\begin{equation}
{\cal L } = \frac{1}{2}( {\bf B}^2 - {\bf E}^2) + \frac{3
\gamma}{2} \, ({\bf E} \cdot {\bf B} )^2 \, . \label{kern-eq.2.6}
\end{equation}
The equations of motion also display a clear physical meaning
when expressed in terms of $\bf E$ and $\bf B$. The equation
(\ref{kern-eq.2.5}) becomes
\begin{equation}
{\bf div}\,{\bf B} = 0 , \, \ \ \, \ \ \, \ \ {\bf rot} \, {\bf
E} = -
\partial_0 {\bf B} \, ,
\label{kern-eq.2.7}
\end{equation}
whereas the equations (\ref{kern.eq.2.4}) become
\begin{eqnarray}
& {\bf div} \, {\bf E} = - 3 \gamma \, {\bf B} \cdot {\bf grad}
\, ( {\bf E} \cdot {\bf B}) &\nonumber\\& {\bf rot} \, {\bf B}
= \partial_0 {\bf E} + 3 \, \gamma \, \biggl[ \, {\bf B}
\partial_0 ({\bf E} \cdot {\bf B}) - {\bf E} \times {\bf grad}
( {\bf E} \cdot {\bf B}) \biggr] \, , \label{kern-eq.2.8}
\end{eqnarray}
which show how the density of charge and the current are
created by the non-linearity of the field: indeed, we can
introduce
\begin{equation}
\rho = - 3 \gamma \, {\bf B} \cdot {\bf grad} \, ( {\bf E} \cdot
{\bf B}) \, \ \ \, \ \ {\rm and} \, \ \ \, \ \ \, {\bf j} = 3
\, \gamma \, \biggl[ \, {\bf B}
\partial_0 ({\bf E} \cdot {\bf B})
- {\bf E} \times {\bf grad} ( {\bf E} \cdot {\bf B}) \biggr] 
\label{kern-eq.2.9}
\end{equation}
which satisfy the continuity equation
\begin{equation}
\partial_0 \rho + {\bf div} \, {\bf j} = 0 \, .
\label{kern-eq.2.10}
\end{equation}
The Poynting vector conserves its form known from the
Maxwellian theory, but the energy density is modified:
\begin{equation}
{\bf S} = {\bf E} \times {\bf B} , \, \ \ \, \ \ \, {\cal{E}} =
\frac{1}{2} \, ({\bf E}^2 + {\bf B}^2) + \frac{3 \gamma}{2} \,
({\bf E} \cdot {\bf B})^2 \, , \label{kern-eq.2.11}
\end{equation}
with the continuity equation resuming the energy conservation
satisfied by virtue of the equations of motion:
\begin{equation}
\partial_0 {\cal E} + {\bf div} {\bf S} = 0 \, .
\label{kern-eq.2.12}
\end{equation}
It can be easily proved that there is birefringence in this
theory. One wave propagates in a Maxwellian way, the other
possible wave solution propagates differently; in fact, it is
delayed (see \cite{LemosKerner} for details).
\newline
\indent The properties of possible stationary axisymmetric
solutions, endowed with non-vanishing charge, intrinsic kinetic
and magnetic moments, have been discussed in
\cite{Kerner1,Kerner2}. In the theory based on the Gauss-Bonnet
term in $5$ dimensions, one can try to find axially-symmetric
configurations displaying both finite electric charge and
finite magnetic moment; also, a kinetic momentum can be
expected, parallel to the magnetic moment.
\newline
\indent In cylindric coordinates $\rho, \varphi, z$ we expect the
induced current density to be aligned on the ${\bf
e}_{\varphi}$-vector of the local frame, giving a current density
circulating around the $z$-axis; the fields ${\bf E}$ and ${\bf
B}$ should be contained in the $\rho-z$ planes orthogonal to
${\bf e}_{\varphi}$. Recalling the fact that the lines of strength
of ${\bf B}$ must be closed, the best description of this
configuration can be obtained using the toroidal curvilinear
coordinates $(\mu, \eta, \varphi)$ defined as follows in terms of
cylindric coordinates:
$$ 
\rho = \frac{a \, \cosh \, \mu}{\cosh \, \mu - \cos \, \eta} , \, \ 
\ \,
\qquad
z = \frac{a \, \sin \, \eta }{\cosh \, \mu - \cos \, \eta}, \,
\ \qquad  \varphi . 
$$ 
with $0 \leq \varphi < 2 \pi \, , \, \ \ 0 \leq \eta
< 2 \pi \, , \, \ \ 0 \leq \mu \leq \infty .$ The coordinate
lines of $\varphi$ are concentric circles in the $(z = 0)$-plane,
while the coordinate lines of the variable $\eta$ are excentric
tori concentrating around the circle $\rho = a$. We shall
suppose that the lines of force of the magnetic field coincide
with the coordinate curves given by $\varphi = $ Const. and $\mu
=$ Const. The configuration we seek can be written as:
\begin{eqnarray}
{\bf E} = E_{\rho} \, {\bf e}_{\rho} +  E_z \, {\bf e}_z =
 E_{\mu} {\bf e}_{\mu} +  E_{\eta} \, {\bf e}_{\eta} ; \\
{\bf B} = B_{\rho} \, {\bf e}_{\rho} +  B_z \, {\bf e}_z =
B_{\eta} \, {\bf e}_{\eta} ; \, \ \ \, \ \ {\bf j}_{ind} =
j_{ind} \, {\bf e}_{\varphi} . \label{kern-eq.3.13}
\end{eqnarray}
It can be also shown that the whole problem can be reduced to
determining just two unknown functions of the variables $(\mu,
\eta)$, because $ {\bf B} = rot {\bf A} $ and $B_{\mu} = 0$,
and because here ${\bf E} = - grad V$, we have $ {\bf A} =
A(\mu, \eta) \, {\bf e}_{\varphi}$, and $V= V(\mu, \eta)$.
\newline
\indent Approximate solutions of this form have been found in
(\cite{Kerner1}, \cite{Kerner2}); here we shall only remind
their essential features. At great distances, the fields ${\bf
E}$ and ${\bf B}$ behave as if they were generated by a finite
charge $Q$ and a finite magnetic dipole ${\bf m}$:
\begin{equation}
{\bf E}_{\infty} \simeq \frac{Q \, {\bf r}}{4 \pi r^3} \, , \ \
\, \ \ {\bf B}_{\infty} \simeq \frac{{\bf m} \wedge {\bf r}}{4
\pi r^3}  \, . \label{kern-eq.3.14}
\end{equation}
The charge is concentrated around the circle $\rho = a$ and
"smeared" in its vicinity; if it is chosen to be positive,
there is a little "halo" of negative charge density farther
away, imitating the vacuum polarization effect. The charge
density's fall-off is vary rapid, behaving at short distances
as $r^{-9}$, and then falling off exponentially; the same
concerns the density of induced current ${\bf j}_{ind}$ which
falls off as $r^{-8}$. The induced current behaves as if it
were produced by the charge density rotating around the
$z$-axis with the speed of light.
\newline
\indent Another interesting feature of this solution is its
$Z_2 \times Z_2$ symmetry. Indeed, any such solution displaying
the total energy (mass) ${\cal{E}}$, the total charge $Q$,
magnetic momentum ${\bf m}$ and the total spin ${\bf{s}}$
is followed by three similar solutions with the same energy,
but either with the same charge, but with the spin and magnetic
momentum in the opposite direction (both ``down''), or another
couple of solutions having the opposite charge, and spin and
magnetic moment up or down, but always opposite to each other -
just like with what we know about the electron and the
positron. The following table shows the properties of the four
solutions:

\begin{center}
\begin{tabular}{|l|l|l|l|l|}  \hline
$Fields$      &         $Energy$    & $Charge$  
& \quad ${\bf m}$ 
& \quad $Spin$   
\\ 
\hline   $\phantom{-}{\bf E}$, $\phantom{-}{\bf B}$     
& $\quad {\cal{E}} $ &\quad $\phantom{-}Q$ & \quad $\phantom{-}{\bf m}$
& \quad $\phantom{-}{\bf s}$ \quad
\\
  $\phantom{-}{\bf E}$,  $-{\bf B}$ 
  & $\quad {\cal{E}}$ &  \quad $\phantom{-}Q$
&\quad
$-{\bf m}$ & \quad $-{\bf s}$ \quad
\\
$-{\bf E}$,   $\phantom{-}{\bf B}$   & $\quad {\cal{E}}$ &  \quad $-Q$
&\quad
$\phantom{-}{\bf m}$ & \quad $-{\bf s}$ \quad
\\
 $-{\bf E}$,  $-{\bf B}$   & $\quad {\cal{E}}$ &  \quad $-Q$
&\quad
$- {\bf m}$ & \quad $\phantom{-}{\bf s}$ \quad
\\
\hline
\end{tabular}
\vglue 0.3cm Tab 1. The symmetry properties of four solutions.
\end{center}
\vskip 0.2cm \indent Unfortunately, these solutions present a
mild singularity on the circle $\rho = a$, which can not be
avoided. Its presence can be proved by using Poincar\'e's
lemma; the details can be found in (\cite{Kerner1},
\cite{Kerner2}).

\section{An SU(2)-Based Non-Abelian Generalization of
 Born-Infeld Theory} 

The superstring theory
gives rise to one important modification of the standard
Yang-Mills quadratic Lagrangian suggesting the action of the
Born-Infeld (BI) type \cite{Ts97,GaGoTo98,BrPe98}. Because this
modification breaks the scale invariance, the natural question
arises whether in the Born--Infeld--Yang--Mills (BIYM) theory
the non-existence of classical particle-like solutions can be
overruled. Although a mere scale invariance breaking, being a
necessary condition, by no means guarantees the existence of
particle-like solutions, a detailed study (\cite{Dimark} has
shown that the $SU(2)$ BIYM classical glueballs indeed do exist.
\newline
\indent Non--Abelian generalization of the Born--Infeld action
presents an ambiguity in specifying how the trace over the the
matrix--valued fields is performed in order to define the
Lagrangian \cite{Ts97,GrMoSc99}. Here we adopt the version with
the ordinary trace which leads to a simple closed form for the
action. The BIYM action with the ordinary trace looks like a
straightforward generalisation of the corresponding $U(1)$
action in the ``square root'' form
\begin{equation}
S=\frac{\beta^2}{4\pi}\int\;(1-{\cal{R}})\;d^4x\, ,
\label{kern-eq.3.1}
\end{equation}
where
\begin{equation}
{\cal{R}}=\sqrt{1+\frac{1}{2\beta^2} F^a_{\mu\nu}F_a^{\mu\nu}
-\frac{1}{16\beta^4}(F^a_{\mu\nu}{\tilde F}_a^{\mu\nu})^2}\, .
\label{kern-eq.3.2}
\end{equation}
It is easy to see that the BI non-linearity breaks the
conformal symmetry ensuring the non-zero trace of the
stress--energy tensor
\begin{equation}
T^\mu_\mu={\cal{R}}^{-1}\left[4\beta^2(1-{\cal{R}})-
F^a_{\mu\nu}F_a^{\mu\nu} \right] \neq 0\, . \label{kern-eq.3.3}
\end{equation}
\indent This quantity vanishes in the limit $\beta\to \infty$
when the theory reduces to the standard one. For the Yang-Mills
field we assume the usual monopole ansatz
\begin{equation}
A_0^a=0\, , \quad A_i^a=\epsilon_{aik}{n^k\over r}(1-w(r)) \,,
\label{kern-eq.3.4}
\end{equation}
where  $n^k=x^k/r,\; r=(x^2+y^2+z^2)^{1/2}$, and $w(r)$ is the
real-valued function. After the integration over the sphere in
(\ref{kern-eq.3.1}) one obtains a two-dimensional action from
which  $\beta$ can be eliminated by the coordinate rescaling
$\sqrt{\beta} t\to t,\; \sqrt{\beta} r\to r$. The following
static action results then:
\begin{equation}
S=\int L dr, \quad L=r^2(1-{\cal{R}})\, , \label{kern-eq.3.5}
\end{equation}
with
\begin{equation}
{\cal{R}}=\sqrt{1+ 2\frac{{w'}^2}{r^2} + \frac{(1-w^2)^2}{r^4}}\, ,
\label{kern-eq.3.6}
\end{equation}
where prime denotes the derivative with respect to r. It is
worth noticing that the non-linearity arises here because of
the non-linear dependence of the tensor $F^a_{\mu \nu}$ on the
potentials $A^b_{\mu}$. The corresponding equation of motion
reads
\begin{equation}
\left(\frac{w'}{{\cal{R}}}\right)'=
\frac{w(w^2-1)}{r^2{\cal{R}}}\, . \label{kern-eq.3.7}
\end{equation}
\indent A trivial solution $w\equiv 0$ corresponds to the
 point-like magnetic BI-monopole with the unit magnetic charge
(embedded $U(1)$ solution). In the Born--Infeld theory it has a
finite self-energy \cite{Gi98}. For time-independent
configurations the energy density is equal to minus the
Lagrangian, so the total energy (mass) is given by the integral
\begin{equation}
M=\int_0^\infty ({\cal{R}}-1)r^2 dr\, . \label{kern-eq.3.8}
\end{equation}
For $w\equiv 0$ one finds
\begin{equation}
M = \int \left(\sqrt{r^4+1}-r^2\right)dr =
\frac{\pi^{3/2}}{3{\Gamma (3/4)}^2}\approx 1.23604978 \, .
\label{kern-eq.3.9}
\end{equation}
Looking now for the essentially non--Abelian solutions of
finite mass, we observe that in order to assure the convergence
of the integral (\ref{kern-eq.3.8}) the quantity ${\cal{R}}-1$
must fall down faster than $r^{-3}$ as $r\to \infty$. Thus, far
from the core the BI corrections have to vanish and the
Eq.(\ref{kern-eq.3.7}) should reduce to the ordinary Yang-Mills
equation, equivalent to the following two-dimensional autonomous
system \cite{Chern78,Kerner,Pr79,BrFoMa94}:
\begin{equation}
\dot w=u, \quad \dot u=u+(w^2-1)w\, , \label{kern-eq.3.10}
\end{equation}
where a dot denotes the derivative with respect to $\tau=\ln
r$. This dynamical system has three non-degenerate stationary
points $(u=0, w=0,\pm1)$, from which $u=w=0$ is a focus, while
two others $u=0,\,w=\pm 1$ are saddle points with eigenvalues
$\lambda =-1$ and $\lambda =2$. The separatices along the
directions $\lambda =-1$ start at infinity and after passing
through the saddle points go to the focus with the eigenvalues
$\lambda=(1\pm i\sqrt{3})/2$.
\newline
\indent It has been proved in (\cite{Dimark}) that {\em the
only finite-energy configurations with non-vanishing magnetic
charge are the embedded U(1) BI-monopoles}. Indeed, such
solutions should have asymptotically $w=0$, which does not
correspond to bounded solutions unless $w\equiv 0$. The
remaining possibility is $w=\pm 1, \dot w=0$ asymptotically,
which corresponds to zero magnetic charge. Coming back to
$r$-variable one finds from (\ref{kern-eq.3.7})
\begin{equation}
w=\pm 1+ \frac{c}{r} + O(r^{-2})\, , \label{kern-eq.3.11}
\end{equation}
where $c$ is a free parameter. This gives a convergent integral
(\ref{kern-eq.3.8}) as $r\to\infty$. The two values $w=\pm 1$
correspond to two neighboring topologically distinct Yang-Mills
vacua.
\newline
\indent Now consider local solutions near the origin $r=0$. For
convergence of the total energy (\ref{kern-eq.3.8}), $w$ should
tend to a finite limit as $r\to 0$. Then using the
Eq.(\ref{kern-eq.3.7}) one finds that the only  allowed
limiting values are $w=\pm 1$ again. In view of the symmetry of
(\ref{kern-eq.3.10}) under reflection $w\to \pm w$, one can
take without loss of generality $w(0)=1$. Then the following
Taylor expansion can be checked to satisfy the
Eq.(\ref{kern-eq.3.10}):
\begin{equation}
w=1-br^2+\frac{b^2(44 b^2+3)}{10(4 b^2+1)} r^4 +O(r^6)\, ,
\label{kern-eq.3.12}
\end{equation}
with $b$ being (the only) free parameter.
\newline
\indent As $r\to 0$, the function ${\cal{R}}$ tends to a finite
value
\begin{equation}
{\cal{R}}={\cal{R}}_0+ O(r^2),  \ \ \, \ \ \,  {\cal{R}}_0=1+12
b^2\, , \label{kern-eq.3.15}
\end{equation}
therefore it is not a solution of the initial system
(\ref{kern-eq.3.8}). What remains to be done is to find
appropriate values of constant $b$ leading to smooth
finite-energy solutions by gluing together the two asymptotic
solutions between $0$ and $\infty$.
\newline
\indent It has been proved in (\cite{Dimark} that {\em any
regular solution  of the} Eq.(\ref{kern-eq.3.7}) {\em belongs
to the one-parameter family of local solutions}
(\ref{kern-eq.3.12}) {\em near the origin}.
\newline
\indent It follows that the global finite energy solution
starting with (\ref{kern-eq.3.12}) should meet some solution
from the family  (\ref{kern-eq.3.11}) at infinity. Since both
these local solutions are non--generic, one can at best match
them for some discrete values of parameters. This technique has
been used first in (\cite{Kerner})
\newline
\indent For some precisely tuned value of $b$ the solution will
remain a monotonous function of $\tau$ reaching  the value $-1$
at infinity (Fig.1). This happens for $b_1=12.7463$.
\newline
\indent By a similar reasoning one can show that for another
fine-tuned value $b_2>b_1$ the integral curve $w(\tau)$ which
has a minimum in the lower part of the strip and then becomes
positive will be stabilized by the friction term in the upper
half of the strip $[-1, \, 1]$ and tend to $w=1$. This solution
will have two nodes. Continuing this process we obtain the
increasing sequence of parameter values $b_n$ for which the
solutions remain entirely within the strip $[-1,\,1]$ tending
asymptotically to $(-1)^n$. The lower values $b_n$ found
numerically are given in Tab. 2.

\begin{center}
\begin{tabular}{|l|l|l|l|}  \hline
$n$           & $\quad b$            & $\quad M   $         \\
\hline
$1$           & $\quad 1.27463 \times 10^1\quad $ &\quad $1.13559$ =
\quad\\
$2$           & $\quad 8.87397 \times 10^2$ &  $\quad 1.21424$ \quad\\
$3$           & $\quad 1.87079 \times 10^4$ &  $\quad 1.23281$ \quad\\
$4$           & $\quad 1.27455 \times 10^6$ &  $\quad 1.23547$ \quad\\
$5$           & $\quad 2.65030 \times 10^7$ &  $\quad 1.23595$ \quad\\
\hline
\end{tabular}
\vglue 0.4cm Tab 2. Parameters $b, \,M$  for first five
solutions.
\end{center}

\section{An SU(2) $\times$ U(1) Generalization of
 Born-Infeld Lagrangian 
 and its Embedding in the Standard Electroweak Model} 

The Born-Infeld Lagrangian generalizes the usual
Maxwell theory; however, since we know that this theory is a
part of the non-abelian field theory which accounts for
electromagnetic and weak interactions, a natural question can
be asked: is the original abelian version of Born-Infeld theory
just a ``shadow'' of a more complicated non-abelian analog of
the Born-Infeld Lagrangian ? If so, we should be able to
compare the pure electromagnetic (abelian) BI-Lagrangian with
what can be extracted from its non-abelian version based on the
symmetry group $SU(2)\times U(1)$ after defining physical
fields as linear combinations of the $U(1)$ and $SU(2)$ gauge
fields with the coefficients defined by a rotation with the
Weinberg angle. The ultimate comparison is beyond the scope of
this paper; we will show how the first few terms in the Taylor
expansions of these two Lagrangians can be compared. Performing
the expansion of the BI Lagrangian and we obtain the following
first few terms in the series up to the fourth order.
\begin{align}                                                                       
 L_{BI}\!&=
-\frac{1}{4} F_{\mu \nu }F^{\mu \nu                                                    
}+\frac{1}{32}\beta^{-2}                                                               
(F_{\mu \nu }F^{\mu \nu })^{2}+\frac{1}{32}\beta^{-2}( F_{\mu \nu }%
\widetilde{F}^{\mu \nu })^{2}\cr &\, -  \frac{1}{128}\beta^{-4}
(F_{\mu \nu }F^{\mu \nu                       
})^{3}-\frac{1}{128} \beta^{-4} F_{\mu \nu }F^{\mu \nu }(F_{\mu                        
\nu }\widetilde{F}^{\mu \nu })^{2}                                                     
+ \frac{5}{2048}\beta^{-6}
(F_{\mu \nu }F^{\mu \nu })^{4}\cr &\, + \frac{3}{1024}
 \beta^{-6}(F_{\mu \nu }F^{\mu \nu         
})^{2}(F_{\mu \nu }\widetilde{F}^{\mu \nu                                              
})^{2}+\frac{1}{2048}\beta^{-6} (F_{\mu \nu }\widetilde{F}^{\mu                        
\nu })^{4}\, .                                                                         
\label{kern-eq.4.1}                                                                    
\end{align}                     
For non-abelian groups we shall use the same generalization of
the Born-Infeld Lagrangian as in the previous section,
(\ref{kern-eq.3.1}). Here we will construct the non-abelian
Lagrangian for $SU(2)$ and $U(1)$, to be compared with
(\ref{kern-eq.4.1}).  We expand the series in powers of
$\beta^{-2}$ in terms of Lorentz invariants of the fields, the
abelian ones, $P $ and $S $, and their non-abelian
generalizations $P'$ and $S'$:
\begin{equation}
P'\equiv F_{\mu \nu }F^{\mu \nu }\, ,\medskip  \, \ \ \, \ \ \,
S'\equiv F_{\mu \nu }\widetilde{F}^{\mu \nu
}=\frac{1}{2}\epsilon ^{\mu \nu \rho \sigma }F_{\mu \nu
}F_{\rho \sigma }\, , \label{kern-eq.4.2}
\end{equation}
with $F_{\mu \nu }=F_{\hspace{0.1cm}\mu \nu }^{a}J_{a}$, with $a=0$ =
for $%
U(1) $ and $a=1..3$ for $SU(2)$. With the invariants
(\ref{kern-eq.4.2}) replacing the abelian ones, and taking all
the traces in the Lagrangian, which in the non-abelian case
takes value in the matrix algebra of the fundamental
representation of $SU(2) \times U(1)$ chosen here, we obtain
\begin{equation}
\begin{array}{c}
L_{SU(2)U(1)}=2 a\hspace{0.05cm}
\hspace{0.05cm}F_{\hspace{0.1cm}\mu \nu }^{0}
F_{\hspace{0.1cm}}^{0\mu \nu } + \frac{1}{2}a\hspace{0.05cm}
\hspace{0.05cm}F_{\hspace{0.1cm}\mu \nu }^{a}
F_{\hspace{0.1cm}}^{a\mu \nu } \medskip \\
+ \beta^{-2}M^{2}\{b\hspace{0.05cm} \hspace{0.05cm}
[2F_{\hspace{0.1cm}\mu \nu }^{0}F_{\hspace{0.1cm}}^{0\mu \nu
}F_{\hspace{0.1cm}\rho \sigma }^{0} F_{\hspace{0.1cm}}^{0\rho
\sigma }+\frac{1}{8} F_{\hspace{0.1cm}\mu \nu
}^{a}F_{\hspace{0.1cm}}^{a\mu \nu } F_{\hspace{0.1cm}\rho
\sigma }^{c}\medskip
F_{\hspace{0.1cm}}^{c\rho \sigma } \\
+ F_{\hspace{0.1cm}\mu \nu }^{0}F_{\hspace{0.1cm}}^{0\mu \nu }
F_{\hspace{0.1cm}\rho \sigma }^{a}F_{\hspace{0.1cm}}^{a\rho
\sigma } +2F_{\hspace{0.1cm}\mu \nu
}^{0}F_{\hspace{0.1cm}}^{a\mu \nu } F_{\hspace{0.1cm}\rho
\sigma }^{0}F_{\hspace{0.1cm}}^{a\rho \sigma }] +
c\hspace{0.05cm} \hspace{0.05cm}[2F_{\hspace{0.1cm}\mu \nu }^{0}
\widetilde{F}\medskip _{\hspace{0.1cm}}^{0\mu \nu }
F_{\hspace{0.1cm}\rho \sigma }^{0}
\widetilde{F}_{\hspace{0.1cm}}^{0\rho \sigma } \\
+\frac{1}{8}F_{\hspace{0.1cm}\mu \nu }^{a}
\widetilde{F}_{\hspace{0.1cm}}^{a\mu \nu }
F_{\hspace{0.1cm}\rho \sigma }^{c}
\widetilde{F}_{\hspace{0.1cm}}^{c\rho \sigma }
+F_{\hspace{0.1cm}\mu \nu }^{0}
\widetilde{F}_{\hspace{0.1cm}}^{0\mu \nu }
F_{\hspace{0.1cm}\rho \sigma }^{a}
\widetilde{F}_{\hspace{0.1cm}}^{a\rho \sigma }
+2F_{\hspace{0.1cm}\mu \nu }^{0}
\widetilde{F}_{\hspace{0.1cm}}^{a\mu \nu }
\medskip F_{\hspace{0.1cm}\rho \sigma }^{0}
\widetilde{F}_{\hspace{0.1cm}}^{a\rho \sigma }]\} +... \, ,
\end{array}
\label{kern-eq.4.3}
\end{equation}
where $\beta$ is the BI-Lagrangian parameter, $M$ the mass
scale of the unified theory, and $a, \, b , \, c, \,...$ are
complicated numerical coefficients coming from traces and
representation-dependent. Introducing physical fields with
linear combinations of the $U(1)$ and the $SU(2)$ gauge fields
\begin{eqnarray}
F_{\hspace{0.1cm}\mu \nu }^{0} &= & 
F_{\mu \nu }\cos \theta
-(\partial _{\mu }Z_{\nu }-\partial _{\nu }Z_{\mu })\sin \theta
\, ,\medskip  \label{l1}
\\
F_{\hspace{0.1cm}\mu \nu }^{3} &= & 
F_{\mu \nu }\sin \theta
+(\partial _{\mu
}Z_{\nu }-\partial _{\nu }Z_{\mu })\cos \theta \medskip =
+ig(W_{\hspace{0.05cm%
}\hspace{0.05cm}\mu }^{\dagger
}W_{\hspace{0.05cm}\hspace{0.05cm}\nu
}^{+}-W_{\hspace{0.05cm}\hspace{0.05cm}\mu}
^{+}W_{\hspace{0.05cm}\hspace{%
0.05cm}\nu }^{\dagger })\, ,  \label{l2}
\\
F_{\hspace{0.1cm}\mu \nu }^{1}
 &= & \frac{1}{\sqrt{2}}[(\partial
_{\mu } W_{\nu}-\partial _{\nu }W_{\mu })+(\partial _{\mu
}W_{\hspace{0.05cm}
\hspace{0.05cm%
}\nu }^{\dagger }-\partial _{\nu
}W_{\hspace{0.05cm}\hspace{0.05cm}\mu
}^{\dagger })\medskip ] 
\nonumber \\
&&- \ 
ig[(W_{\nu }-W_{\hspace{0.05cm}\hspace{0.05cm}\nu }^{\dagger })
(A_{\mu}\sin \theta +Z_{\mu }\cos \theta ) 
\nonumber \\
&&+ \ 
 (W_{\mu }-W_{\hspace{0.05cm}\hspace{0.05cm}\mu }^{\dagger })
(A_{\nu }\sin \theta +Z_{\nu }\cos \theta )] \, ,
\label{kern-eq.4.4}
\end{eqnarray}
\vbox{
\begin{eqnarray}
F_{\hspace{0.1cm}\mu \nu }^{2} &= & 
\frac{i}{\sqrt{2}}[(\partial
_{\mu }W_{\nu
}-\partial _{\nu }W_{\mu })-(\partial _{\mu}
W_{\hspace{0.05cm}\hspace{0.05cm%
}\nu }^{\dagger }-\partial _{\nu
}W_{\hspace{0.05cm}\hspace{0.05cm}\mu
}^{\dagger })\medskip ] 
\nonumber \\
&&+ \ g[(W_{\nu }+W_{\hspace{0.05cm}\hspace{0.05cm}\nu }^{\dagger
})(A_{\mu }\sin
\theta +Z_{\mu }\cos \theta ) 
\nonumber \\
&&- \ 
(W_{\mu }+W_{\hspace{0.05cm}\hspace{0.05cm}\mu }^{\dagger
})(A_{\nu }\sin \theta +Z_{\nu }\cos \theta )] \, ,
\label{kern-eq.4.5}
\end{eqnarray}
}
where $A_{\mu }$ is the pure electromagnetic field, $Z_{\mu }$
is the neutral boson, $W_{\hspace{0.05cm}\hspace{0.05cm} \mu
}^{+}\ $ and $W_{\hspace{0.05cm}\hspace{0.05cm}\mu }^{-}$ are
the charged $W$-bosons, we can now compare the two series, term
by term, trying to fix the coefficients in order to make
coincide as many terms as possible. With the Weinberg angle
$\theta $ we can identify the pure electromagnetic sector in
(\ref{kern-eq.4.3}), then evaluate the difference. Because of
the lack of space, we show here only first terms of this
expression:
\begin{equation}
\begin{array}{c}
L_{SU(2)U(1)}-L_{EM}=a (2\cos^{2} \theta +\frac{1}{2} \sin^{2}
\theta ) F_{\mu \nu }
F_{\hspace{0.1cm}}^{\mu \nu } \medskip \\
+\beta^{-2}M^{2}\{b [2\cos ^{4}\theta +\frac{1}{8}\sin
^{4}\theta +3\sin ^{2}\theta \cos^{2}\theta ] F_{\mu \nu }
F_{\hspace{0.1cm}}^{\mu \nu } F\medskip _{\rho \sigma}
F_{\hspace{0.1cm}}^{\rho \sigma } \medskip \\
+  c [2\cos ^{4}\theta +\frac{1}{8}\sin ^{4}\theta +3\sin
^{2}\theta \cos^{2}\theta ] F_{\mu \nu
}\widetilde{F}_{\hspace{0.1cm}}^{\mu \nu }F_{\rho \sigma }
\widetilde{F}\medskip _{\hspace{0.1cm}}^{\rho \sigma }] \} \medskip \\
+\beta^{-4}M^{3}\{g\hspace{0.05cm} \hspace{0.05cm}[\frac{1}{32}
\sin ^{6}\theta +\frac{15}{8}\cos ^{2}\theta \sin ^{4}\theta
+\frac{15}{2}\cos ^{4}\theta \medskip \sin ^{2}\theta
+2\cos^{6}\theta ] + ...\} \, .
\end{array}
\label{kern-eq.4.6}
\end{equation}
It is possible to show that the coefficient for the $n$-th
order of $\beta^{-2}$ is given by
\begin{equation}
C_{n}\left( \theta \right) =\frac{1}{4^{n}}[(1-2\cot \theta
)^{2n}+(1+2\cot \theta )^{2n}](\sin \theta )^{2n}\, ,
\label{kern-eq.4.7}
\end{equation}
and its derivative is given by
\begin{eqnarray}
C_{n}^{\prime }(\theta )&= &\frac{n}{2^{2n-1}}\{\cot \theta
\lbrack (1-2\cot
\theta )^{2n}+(1+2\cot \theta )^{2n}] \nonumber \\
&&+ \, 2[(1-2\cot \theta )^{2n-1}+(1+2\cot \theta )^{2n-1}](\csc
\theta )^{2}\}(\sin \theta )^{2n}\, .
\label{kern-eq.4.8}
\end{eqnarray}
\indent It is interesting to examine the behaviour of these
coefficients. Surprisingly enough, starting from $n=3$ they
display a maximum, whose position converges to a certain value
with growing $n$; moreover, this position is very close to the
established value of the Weinberg angle ( satisfying $\sin
^{2}\theta _{W}=0.227\pm 0.014$, corresponding to $\theta_W
=28^o,45$ or to $0.497$ radians). The maxima were found solving
$C_{n}^{\prime }(\theta )=0$ for a given value of $n$. We show
below examples of the coefficients $C_{n}$ starting from $n=3$,
and the value of the angle (in radians) for the first maximum
of $C_{n}$

\begin{equation}
C_{3}=\frac{1}{32}\sin ^{6}\theta +\frac{15}{8}\sin ^{4}\theta
\cos ^{2}\theta +\frac{15}{2}\sin ^{2}\theta \cos ^{4}\theta
+2\cos ^{6}\theta\, . \label{kern-eq.4.9}
\end{equation}
The first maximum corresponds to $C_{3}=2.083\,8$ for $\theta
=0.346\,82.$

Later on, we have:

First maximum for $C_{4}=2.4886$ at $\theta =0.\allowbreak
43522.$

First maximum at $C_{5}=3.071\,3\ $for $\theta =0.\,\allowbreak
45474.$

First maximum at $C_{6}=3.823\,2$ for $\theta =0.4606$.

For $n=8$ we obtain

First maximum at $C_{8}=5.962\,2$ for $\theta =0.463\,27$.

The value of the angle corresponding to the first maximum for
higher order tends to $\theta =0.463648$ and remains constant
for  $n=50$ and higher.

The fact that the value of the mixing angle obtained for the
first maximum of the of the so defined coefficients approaches
the value of the Weinberg angle seems to be rather accidental;
nevertheless, it is worth noticing.

\section*{Acknowledgments}
A. L. Barbosa thanks FAPESP (S\~{a}o Paulo, Brazil) for
financial support. D. V. Gal'tsov  acknowledges support from
RFBR under grant 00-02-16306.
                                                                                                       
\end{document}